\def\etal {{\rm et al.}\ }

\def\littlesm{\ifmmode{\scriptscriptstyle m }
    \else{\hbox{$\scriptscriptstyle m $ }}\fi}
\def\littleprime{\ifmmode{\scriptscriptstyle \prime }
    \else{\hbox{$\scriptscriptstyle \prime$ }}\fi}
\def\littlecirc{\ifmmode{\scriptscriptstyle \circ }
    \else{\hbox{$\scriptscriptstyle \circ $ }}\fi}
\def\littless{\ifmmode{\scriptscriptstyle s }
    \else{\hbox{$\scriptscriptstyle s $ }}\fi}
\def\magm{\raise .9ex \hbox{\hskip-1pt\littlesm}}
\def\arcsec{\raise .9ex \hbox{\littleprime\hskip-3pt\littleprime}}
\def\arcmin{\raise .9ex \hbox{\littleprime}}
\def\degree{\raise .9ex \hbox{\littlecirc}}
\def\magpoint{\hbox to 1pt{}\rlap{\magm}.\hbox to 2pt{}}
\def\arcsecpoint{\hbox to 1pt{}\rlap{\arcsec}.\hbox to 2pt{}}
\def\arcminpoint{\hbox to 1pt{}\rlap{\arcmin}.\hbox to 2pt{}}
\def\degreepoint{\hbox to 1pt{}\rlap{\degree}.\hbox to 2pt{}}

\documentstyle[12pt]{report}
\begin{document}

\centerline {\bf Abstract}

We have obtained the radii and distances of 16 galactic Cepheids supposed
to be members in open clusters or associations using the new optical and
near-infrared calibrations of the surface brightness (Barnes-Evans) method
given by Fouqu\'e \& Gieren (1997).  We find excellent agreement of the radii
and distances produced by both infrared techniques which use the V, V-K (K
on the Carter system) and
K, J-K magnitude-color combinations, respectively, and typical random
errors as small as $\sim$2 percent. We discuss in detail possible systematic
errors in our infrared solutions and conclude that the typical total
uncertainty of the infrared distance and radius of a Cepheid is about 3
percent in both infrared solutions, provided that the data are of excellent
quality and that the amplitude of the color curve used in the solution is
larger than $\sim$0.3 mag.  The optical V, V-R distance and radius of a given
Cepheid can deviate by as much as $\sim$30 percent from the infrared value, 
due to large systematic and random errors caused by microturbulence and 
gravity variations which affect the optical V-R color, but not the V-K and J-K
colors, as shown by Laney \& Stobie (1995).

We find excellent agreement of our infrared radii with the infrared radii
derived by Laney \& Stobie (1995) for these variables from an application of
the maximum likelihood technique, which further increases our confidence
that the total errors in our infrared solutions are not larger than $\sim$3
percent. In an Appendix we discuss the relative advantages and disadvantages
of our infrared surface brightness technique and the maximum likelihood
technique.
We compare the adopted infrared distances of the Cepheid variables to the
ZAMS-fitting distances of their supposed host clusters and associations and
find an unweighted mean value of the distance ratio of 1.02 $\pm$ 0.04.  A
detailed discussion of the individual Cepheids shows that the uncertainty
of the ZAMS-fitting distances varies considerably from cluster to cluster.
We find clear evidence that four Cepheids are not cluster members (SZ Tau,
T Mon, U Car and SV Vul) while we confirm cluster membership for V Cen and
BB Sgr for which the former evidence for cluster membership was only weak.
After rejection of non-members, we find a weighted mean distance ratio of
0.969 $\pm$ 0.014, with a standard deviation of 0.05, which demonstrates that
both distance indicators are accurate to better than 5\%, including
systematic errors, and that there is excellent agreement between both
distance scales.

\vspace{0.8cm}

\noindent
{\it Subject headings:} Stars: Cepheids - stars: distances - infrared: stars -
  distance scale - \\
 \phantom{Subject headings:} clusters: open

\vspace{1.2cm}

\centerline {\bf 1. Introduction}

Cepheid variables are crucial for the establishment of absolute distances
to a sample of galaxies in the near field ($\sim \leq$ 30 Mpc).  These 
Cepheid-calibrated galaxies will then in turn serve
to calibrate useful secondary distance indicators, like Supernovae Ia (Saha
et al.  1996) which reach out to the far field of the Hubble flow, and from
the observation of these objects in remote galaxies a reliable and accurate
value of the Hubble constant will eventually be derived.

While HST makes it now possible to obtain distances to Virgo cluster
galaxies based on their Cepheid variables which are accurate to 0.15 - 0.20
mag (Ferrarese et al.  1996), these are distances as measured {\it relative
to the LMC}.  The largest uncertainty in deriving the {\it absolute
distances} to these galaxies still lies in our limited ability to measure
the distances to local Cepheid variables, and in the corresponding
uncertainty of the true distance to the LMC.  In view of the enormous
importance of the Cepheid process to establish a basic set of accurate
distances to nearby galaxies, the improvement of the {\it local calibration
of Cepheid distances} is more important than ever.

One of the fundamental techniques to measure the distance, and the radius
of a Cepheid is the Barnes-Evans method (hereafter BE) which makes use of the
observed variations in light, color and radial velocity of the variable.
In this method, it is assumed that the surface brightness
of the variable, at any given phase, can be accurately inferred from a
measured color index at this phase - the degree to which this basic
assumption is fulfilled critically determines both random and systematic
errors in the radius and distance result.  While in the past most applications
of the Barnes-Evans method to Cepheid variables have been carried out  
in the optical spectral range,
more recent work using near-infrared magnitudes and colors (e.g. Welch 1994) 
has shown that
infrared colors are much better surface brightness indicators than optical
ones.  This has been demonstrated particularly convincingly in the work of
Laney \& Stobie (1995) who were able to trace down the physical causes for
the inferiority of optical colors, which are the variable microturbulence
and gravity during a Cepheid's pulsation cycle which invalidate, to a
significant extent, the basic assumption of the Barnes-Evans method for optical
colors while their effect on infrared colors is almost negligible.  A
further, important advantage in going to infrared colors is to avoid
photometric contamination of Cepheid light and color curves by a companion
- most Cepheid companions are blue stars which do affect optical, but not
infrared colors.

Motivated by these results, Fouqu\'e \& Gieren (1997; hereafter Paper I)
have provided new infrared calibrations of Barnes-Evans technique.  In
addition to using the V-K color index which was already used in
a former study of Welch (1994), Paper I also provides a K, J-K version of
the technique which makes only use of near-infrared magnitudes and colors.
One particular strength of the new calibrations given in Paper I is that
the zero point of the surface brightness- color relations was derived
solely on the basis of accurate, interferometrically measured angular
diameters of cool giant and supergiant stars, making the distance results
independent of Cepheid model atmospheres or effective temperature scales
which had to be used in the past.  A first application of the method to the
cluster Cepheid U Sgr had shown that there is a {\it dramatic increase in
accuracy of the infrared solutions}, as compared to the optical one.

The purpose of this second paper is twofold.  Firstly, we shall provide and
discuss near-infrared radii and distances for many more Cepheid variables,
and compare them to the optical solutions which we shall also derive.  This
will provide a clear picture regarding the relative strengths and
weaknesses of the different variants of our technique.  We shall also give a
thorough discussion of the sources and magnitudes of the errors in our
results, and we shall demonstrate that the near-infrared technique is able
to produce radii and distances of Cepheids with a typical accuracy of
$\pm$3 percent.  Secondly, we shall compare the infrared distances to the
distance scale set by cluster ZAMS-fitting - to this end, we are doing our
analyses in this paper on all Cepheids in open clusters and associations
for which the necessary data are available in the literature.  We shall
discuss the Cepheid period-radius and most importantly, period-luminosity
relation (in different passbands) in a forthcoming paper which will
incorporate the solutions on a larger number of variables for an improved
statistics of the results.

Our paper is organized as follows:  in section 2, we present and discuss
the radius and distance solutions; in section 3, we compare the infrared
with the optical results; in section 4, we test the stability of our
solutions to various factors and present an error budget; and in section 5,
we compare our infrared distances to the cluster ZAMS-fitting distances.
In an Appendix, we further discuss the relative merits of our technique as
compared to the maximum likelihood approach used by Laney \& Stobie (1995).

\vspace{1.5cm}

\centerline {\bf 2. Radius and Distance Solutions}

\vspace{0.8cm}

\centerline {\bf 2.1. The Sample of Galactic Cluster Cepheids and adopted Data}

Laney \& Stobie (1992) have obtained extensive near-infrared photometry of
southern Cepheids on the Carter system, including all known or suspected
cluster Cepheids south of $\delta$ = +30$^{o}$, which represent about 75
percent of all galactic Cepheids in open clusters and associations (Feast
\& Walker 1987).  For this sample of cluster Cepheids, we have conducted a
literature survey to determine the highest-quality sources of the other
data needed in our analyses, namely radial velocity curves, V light curves
and V-R color curves.  In this process it became clear that for a few of
the cluster Cepheids some of the needed data sets, mostly the radial
velocity curves, were not of sufficient quality for the analysis of this
paper.Our choice in these (few) cases was to omit these stars in order to
maintain a high and approximately uniform quality in the adopted data for
the selected Cepheids.  This defined a final set of 16 Cepheids adopted for
this study.  The sources of the adopted radial velocity and V,R photometric
observations are given in Table 1.  For all of these Cepheids, the
near-infrared J and K data were taken from Laney \& Stobie (1992), which
ensures homogeneity in the adopted infrared photometry.

An important next step is the choice of the correct pulsation periods for
the variables which ensures a correct phase match between data sets
obtained at different epochs.  Rather than to conduct a new analysis of the
periods, we have relied on the work of Laney \& Stobie (1992) and have
adopted their periods for all variables.  The results in section 2.2 will
show that for most of the Cepheids, the adopted periods lead to excellent
phase matches between data sets spaced by many pulsation cycles, but there
are a few cases where the period values might be slightly improved, or
where the period itself is variable (see section 3.3.).  Another choice we
have to make refers to the color excesses of the variables.  We have chosen
to adopt the excesses given by Fernie (1990) which are usually mean values
from many individual determinations, and have the advantage of having
been derived
on a homogeneous system.  Fortunately, and this being one of the particular
strengths of the BE method, the choice of the adopted color excesses is
not critical for the radius and distance solutions (see section 4.1.).  A
final choice we have to make in our solutions refers to the p-factor which
converts the observed radial into the pulsational velocities of the stellar
surface.  We adopt here the values which come from the formula given in
Gieren, Barnes \& Moffett (1993) which result in values of the p-factor
which depend slightly on period.  There is a possibility that the p-factor
is not constant for a given Cepheid but shows some variation with phase
(Sabbey et al.  1995); our results allow to conclude that such a variation
is probably not important for the BE analysis for most of the Cepheids in
our sample (see section 4.6.).  In Table 2, we provide the values for
periods, reddenings and p-factors we have adopted for this study.  We also
provide information on the uncertainties of the adopted reddenings, which
were calculated as explained in the Appendix.

\vspace{0.8cm}

\centerline {\bf 2.2.  Optical and Near-Infrared Radius and Distance Solutions}

In Paper I, we have derived expressions for the angular diameter of a
Cepheid variable in terms of a de-reddened magnitude and color.  These
expressions are:

$\theta = 10^{0.5474 - 0.2 Vo + 0.760 (V-R)o}$ \hfill (1)

$\theta = 10^{0.5474 - 0.2 Vo + 0.262 (V-K)o}$ \hfill (2)              

$\theta = 10^{0.5474 - 0.2 Ko + 0.220 (J-K)o}$ \hfill (3)

\noindent
Equ. (1) represents the optical calibration in terms of the V, V-R$_{J}$
magnitude/color combination, whereas equ.  (2) and (3) represent the
infrared calibrations in terms of V, V-K and K, J-K.  For each of the Cepheid
variables, we have calculated the angular diameters from each of these
equations using the photometric data from the sources given in Table 1, and
correcting the data sets for possible shifts in phase and photometric zero
point in cases where different data sets were combined.  In the case of the
optical V-R colors, those data sets obtained on the Cousins system were
transformed to the Johnson system following the precepts of Gieren (1984)
and allowing a variable zero point in the transformation formula to achieve
best agreement with the colors observed on the Johnson system in each case.
In order to obtain the V-K color curves of the variables from observations
which in no case were obtained contemporaneously, we adopted the procedure
to fit Fourier series to the K curves from which the values were taken
which corresponded to the phases of the actual V observations.

The angular diameters were then combined with the linear displacements
which were calculated from the integrated radial velocity curves.  In order
to perform the integration, we first fitted Fourier series of appropriate
orders to the observed radial velocity curves of each star, with the
criterion to obtain good fits of the observed structures, without
overfitting the curves, i.e.  introducing artificial undulation in the
fitted curves.  The linear displacements were then calculated at the phases
of the photometric observations.  The radii and distances were obtained
from least-squares solutions of the relation

     $D_{0} + \Delta D  =  10^{-3}~d~\theta$   \hfill               (4)

\noindent
where $D_{0}$ is the mean linear diameter of the Cepheid, $\Delta D$ the
displacement from the mean (both in AU), d the distance (in pc), and
$\theta$ the angular diameter at this particular phase (in mas).  Before
doing the solutions, we plotted the linear and angular diameter curves of
each of the Cepheids to check for the correct phase match of both curves.
In most cases, the curves matched very well and there was no need to apply
a shift to the curves to bring them into phase agreement, but in some cases
(and particularly in the V, V-R solutions) a small phase shift was applied.

In Table 3, we give the radius and distance results we obtained for each
Cepheid in each of the three magnitude/color combinations, together with
the corresponding standard deviations.  A discussion of the errors involved
in the determination of the linear displacements and angular diameters (see
Appendix) shows that in all cases the angular diameters bear larger
relative uncertainties than the linear displacements; under these
circumstances, the correct fit to adopt in the least-squares solutions of
equ. (4) is the {\it inverse} fit (see Appendix).  In order to show the
sensitivity of our radius and distance results to the adopted fit, Table 3
also gives the differences of the adopted inverse to the corresponding
direct fit (which assumes all errors in the linear displacements), in units
of the standard deviation $\sigma$ of the inverse solution.  As can be
seen, the difference is negligible in the infrared solutions while it can
be important in the optical solution.  We discuss this further in section
4.3.  Column 7 of Table 3 gives the adopted phase shifts between
displacement and angular diameter curves, and the last column of Table 3
contains information about the number of photometric points used in the
solution. Ideally, the phase shift between the angular diameter and the linear
displacement curve should be the same in the different solutions because the star is
pulsating with a fixed period. In a few cases, however, the best match occurs
for slightly different phase shifts for the same Cepheid. This is probably a
consequence of the observational scatter, but it could also reflect a deeper
problem with the phase shifts. In any case, a slightly different phase shift
as adopted by us for the infrared solutions of some of the stars in Table 3
does not have any important practical effect on the resulting radius and 
distance values because the infrared solutions are very insensitive to the
adopted phase shifts (see discussion in section 4.3.). 

In Figures 1-3 we show, as a typical example of our solutions,
the variations of linear displacements and angular diameters, and the plots
of linear vs.  angular diameter (from which the distance and radius is
derived), for the Cepheid VY Car as obtained from the 3 different
magnitude/color combinations.

The values of the errors in Table 3 confirm the trend seen in Paper I in
the results for U Sgr, namely that the uncertainties of both infrared
solutions are generally much smaller than the one of the V, V-R solution,
especially for the shorter-period stars.  Typical values for the
uncertainties of our radius and distance results are 5-10 percent in V, V-R,
2-3 percent in K, J-K, and 1-2 percent in V, V-K.  In Figure 4, we have
plotted the uncertainties of the radii and distances (normalized to a fixed
number of 30 photometric observations per star) against the amplitudes of
the color curves used in the solutions.  As expected, there is a strong
correlation:  in the K, J-K solutions, there is sharp increase in the errors
if the amplitude of the color curve drops below 0.1 mag.  This is expected
because in this case the variation of the color term in equ.  (3) is in the
order of only 0.01 mag, i.e.  in the order of the photometric
uncertainties.  On the other hand, at an amplitude of only 0.13 mag in J-K
an accuracy of $\pm$4\% can already be reached, and for the long-period
Cepheids which have typical amplitudes of 0.3 mag the accuracy increases to
$\sim$1\% if the data are good enough and the extrema of the color curve are
sampled by the observations.  The situation in the V, V-K
solutions is even more favorable because the amplitudes of Cepheid V-K
color curves are typically 3 times larger than those of their J-K curves;
Figure 4 shows that even for the lowest-amplitude Cepheids of our sample, a
radius and distance accuracy of $\sim$2\% is easily obtainable, if the data
used in the solutions are of adequate quality.  For this reason, we can
trust the V, V-K radius and distance values of the three shortest-period,
low amplitude Cepheids in our sample, EV Sct, SZ Tau and QZ Nor (but see
the notes in section 2.3), while we cannot place much confidence in the
values coming from the other two magnitude/color combinations.

\vspace{0.8cm}

\centerline {\bf 2.3.  Notes on Individual Cepheids}

In this section, we give notes on some of the stars which are important for
our analysis.  \\ 
{\bf EV Sct:}  The real uncertainty of the V-K radius and distance is
probably higher than the error given in Table 3, due to the small number of
K observations (12) available for the analysis. EV Sct is likely to be an
overtone pulsator, but this does not affect its radius and distance derived
from the Barnes-Evans technique. \\
{\bf SZ Tau:}  The period is known to be variable (Szabados 1977) which may
have caused a phase mismatch between the V and K curves, and thus a
systematic error in the V-K solution (see section 4.4). As EV Sct, SZ Tau
is likely to be a first overtone pulsator. \\
{\bf CV Mon:}  There are considerable differences among published V light
curves which are likely to be due to two nearby companion stars at 10" and
14" (Evans \& Udalski 1994) which have or have not been included in the
aperture photometries.  The data used by us in this paper are not affected
by this problem. \\
{\bf BB Sgr:}  This Cepheid is likely to be a spectroscopic binary (Barnes,
Moffett \& Slovak 1988).  We have used for our analysis the radial velocity
obtained by Gieren (1981 a) during only two consecutive pulsation cycles
during which a possible change of the $\gamma$ velocity of the Cepheid must
be negligibly small, thus not affecting our analysis.  \\
{\bf V340 Nor:}  The only published V data for this variable of Eggen
(1983) and Coulson \& Caldwell (1985 b) show unusual scatter for a Cepheid
of this magnitude, affecting the accuracy of our V, V-R and V, V-K solutions.
This is probably due to the position of the variable close to the center of
NGC 6067 which makes accurate aperture photometry difficult.  It would be
very desirable to obtain a modern V light curve of this Cepheid from CCD
observations and PSF fitting techniques.  \\
{\bf T Mon:}  This Cepheid is known to be a spectroscopic binary (Gieren
1989).  To avoid any problem with the variable $\gamma$ velocity of the star, 
we have adopted the pulsational radial velocity curve given by Coulson (1983)
which is freed from the orbital motion effect.  \\
{\bf SV Vul:}  The period of this very long-period Cepheid is changing
(Bersier et al.  1994).  However, we find that the period value adopted in
this paper gives a satisfactory representation of all the radial velocity
data of Bersier et al., and also defines light curves of reasonably low
scatter.  Nevertheless, there is a possibility of a phase mismatch between
V and K, and thus of a systematic error, in the V-K solutions. Also, it
appears likely that for a Cepheid as extended and luminous as SV Vul one
or more of the assumptions of the Barnes-Evans method might fail, affecting
the derived radius and distance in a systematic way. This possibility should
be borne in mind when interpreting the radius and distance data of the most
luminous Cepheids.

\vspace{1.5cm}

\centerline {\bf 3. Comparison of the Near-Infrared to the Optical Solutions}

\vspace{0.8cm}

\centerline {\bf 3.1. Comparison of V, V-K to K, J-K Radii and Distances}

In Figure 5, we have plotted the ratio of the Cepheid radii as derived from
the V, V-K and K, J-K magnitude/color combinations against period.  For most
of the Cepheids, the agreement of the radius values from the two
calibrations is excellent, better than about 4\%.  Omitting the 3
shortest-period Cepheids of our sample whose K, J-K radii cannot be trusted
for the reasons given above, the mean ratio of the radii of the remaining
13 stars is 0.98 $\pm$ 0.012.  There is clearly no trend of this ratio with
period.  We thus conclude that there is no significant systematic
difference between the V, V-K and K, J-K radius solutions.

In Figure 6, we have plotted the distance ratios vs.  period.  The plot
looks very similar to the corresponding radius plot.  The mean ratio of the
distances is 0.97 $\pm$ 0.013, and there is no trend with period either.  The
mean offset of 3 percent between the two solutions may be marginally
significant, in view of the very small uncertainty of this result, but at
this point we prefer to conclude that there is no significant difference
between the distances as derived from the V, V-K and K, J-K versions of the
method, either.  A small difference in the order of 3\% may however turn out
to be real when we have better statistics, which will be the case in Paper
III.

{\it The important conclusion from this comparison is that there is an
excellent level of agreement between individual radii and distances from
both methods, without any significant or systematic (with period) offset
between the results.}  This is an extremely important result which
indicates that the low errors found in the solutions are real and that
systematic uncertainties in both methods are below the $\sim$3\% level (but
see section 4.6).  {\it As a consequence, we adopt as our final radius and
distance values for each Cepheid the weighted mean of the V, V-K and K, J-K
solutions.}  However, there are the following exceptions:  as discussed
above, the K, J-K solutions for the 3 shortest-period,low-amplitude Cepheids
EV Sct, SZ Tau and QZ Nor are not reliable, and we therefore adopt for
these stars the V, V-K solutions.  Furthermore, in the V, V-K solutions for
the 3 long-period Cepheids T Mon, U Car and SV Vul there are small but
systematic deviations in the shapes of the linear displacement and angular
diameter curves, which are {\it not} seen in the K, J-K solutions.  We show
this for U Car in Figure 7.  The most likely cause for this effect is a
small phase mismatch between the V and K curves of these Cepheids, due to
slightly incorrect period values or variable periods, as in the case of SV
Vul, but the effect may also be due to an increasing departure from the basic
assumptions of the Barnes-Evans methods shown by the longest-period stars
which manifests itself in a more pronounced way in the V, V-K than in the
K, J-K solutions.  We therefore feel that it is the best choice to adopt the 
pure
infrared solutions for these stars.  In practice, this is not an important
issue because the two solutions are very similar for U Car and SV Vul
(there is, however, a larger discrepancy for T Mon for which the phase
mismatch may be more serious).

The final, adopted radius and distance values for the Cepheids are given in
columns 3 and 7 of Tab.  4.

\vspace{0.8cm}

\centerline {\bf 3.2.  Comparison of the Infrared to the Optical Solutions}

In Figures 8 and 9 we compare the optical V, V-R$_{J}$ radii and distances
to the adopted values from the infrared solutions.  Two effects can be
immediately seen in these plots:  firstly, both the optical radii and
distances are on average clearly larger than their infrared counterparts,
and secondly, there is a very significant scatter in these plots.  The mean
values for the ratios R(optical) / R (infrared) and d(optical) /
d(infrared) are 1.13 $\pm$ 0.05 and 1.16 $\pm$ 0.06, respectively.
Disregarding the 3 shortest-period Cepheids for which the V, V-R$_{J}$
solutions have very large random errors due to the very small amplitudes of
the V-R color curve, the data in Figs.  8 and 9 demonstrate that for a
given Cepheid with very good observational data, the deviation of its
optical radius and distance from its infrared counterpart can be up to
about 30 percent in both directions.  This means that the effect of
variable microturbulence and gravity on the optical V-R color does not only
introduce large random, but also large systematic errors.  Our impression
on the basis of the few published data is that these systematic deviations
might be mostly correlated with the mean microturbulence of a Cepheid,
which can vary considerably from one Cepheid to another, even for stars of
very similar pulsation period; this hypothesis could be tested with good
spectroscopic microturbulence determinations.

As a conclusion, our solutions confirm the conclusions of Laney \& Stobie
(1995) about the lack of reliability of the optical radii in individual
cases (the behavior of the optical radii of a large sample of Cepheids will
be discussed in our next paper).  This is, of course, the reason why we
adopt the infrared solutions as our final values, which are obviously not
plagued by the problems of the optical solutions.

\vspace{0.8cm}

\centerline {\bf 3.3.  Comparison of the Infrared Radii to published Work}

In columns 5 and 6 of Table 4, we compare our infrared radii to the values
found by Laney \& Stobie (1995) from the corresponding magnitude/color
combination using the maximum likelihood technique, and to the values
obtained by Gieren, Barnes \& Moffett (1989) from their calibration of the
optical V, V-R$_{J}$ surface brightness method.  Since the Laney \& Stobie radii
were obtained using a p-factor of 1.36 for all Cepheids, we have normalized
our radii to this p-factor in the comparison.  Omitting again the 3
shortest-period Cepheids from the comparison (which have uncertain maximum
likelihood radii for the same reason as in our analysis), the mean value of
the ratio R(LS) / R(this paper) is 0.98 $\pm$ 0.03, which means that {\it
the radii from both sources agree within a very small uncertainty.}  In
Figure 10 we have plotted the radius ratios against period, and it can be
seen that there is no significant trend with period.  The most significant
(20\%) deviation for an individual star occurs for CV Mon; this may be
related to the fact that we were able to use, in this study, a much
improved radial velocity curve as compared to the one available to Laney
and Stobie.  If we omit CV Mon in the comparison, the mean ratio becomes
0.965 $\pm$ 0.013, indicating a marginal possibility that the Laney \&
Stobie radii are on average 3\% smaller than our radii, which is still very
close to our result.  We therefore conclude that {\it the fact that our
infrared radii do agree so well with the ones obtained by Laney \& Stobie
who have derived them using quite a different approach (albeit sharing many
of the data we used in our study, in particular the near-infrared
photometry) lends further strong support to the idea that the infrared
radii are very accurate, at the $\pm$ 3\% level.}

The Gieren, Barnes \& Moffett (1989) radii are on average 11 percent larger
than our adopted infrared radii, a number which shows that they basically
do agree with the V, V-R$_{J}$ radii derived in this study.  This seems
surprising given the fact that our new calibration of the optical surface
brightness relation (equ. 26 in Paper I) tends to make the radii smaller.
However, Gieren et al. used {\it direct} least squares fits for most of their
Cepheids which yield smaller Cepheid radii than the inverse fits we have
used in this paper, and both effects seem to cancel, to a first
approximation.

\vspace{1.5cm}

\centerline {\bf 4. Stability of the Radius and Distance Solutions}

In order to establish the influence of various possible sources of error on
our radius and distance results, we have carried out a series of tests
whose results will be described in the following sections.

\vspace{0.8cm}

\centerline {\bf 4.1.  Absorption Corrections}

Due to the structure of equations 1-3 for the Cepheid angular diameter, it
is immediately seen that an error in the color excess used to correct the
observed magnitude and color for interstellar absorption and reddening will
tend to cancel out; for instance, if the value of E(B-V) used in the
calculation is too large, the absorption-corrected magnitude will be too
bright, but the intrinsic color index comes out too blue, and both errors
will tend to cancel.  Our numerical tests show that the sensitivity of the
optical V, V-R and the infrared V, V-K methods to errors in the adopted color
excess are very similar:  a change of $\Delta$ E(B-V) = 0.15, which
corresponds to a change in the adopted value of Av of about 0.5 mag,
produces a $\sim$ 3\% systematic change in the distance and leaves the
radius virtually unchanged.  The K, J-K version of the method is even more
insensitive to the adopted reddening corrections:  the same 0.15 mag change
in E(B-V) produces only a 1.4\% change in the distance while leaving the
radius completely unaffected.

Since the typical uncertainty of the color excesses of the Cepheids given
in Table 2 is about 0.04 mag, any systematic errors in the radii and
distances due to uncertain reddening corrections will be below the 1
percent level, for both the optical and near-infrared versions of the
method. {\it Absorption corrections are thus not a source of concern in the
method.}

\vspace{0.8cm}

\centerline {\bf 4.2.  Direct or Inverse Least-Squares Fits?}

A discussion of the errors in the linear displacements and angular
diameters (see Appendix) shows that the correct fit to adopt in our
solutions is the {\it inverse} least-squares fit.  However, the data in
columns 4 and 6 of Table 3 show that in both near-infrared solutions, the
choice of the fit for those Cepheids with good observational data is
unproblematic:  the difference between the extreme solutions (direct and
inverse fits) is always smaller than one standard deviation, in these
cases, so it {\it doesn't matter how the fit is taken in the near-infrared
solutions.} This is, of course, another way of saying that the random
errors in the infrared solutions are very small.  A consequence of this is
that we are not dependent on having photometric data covering the complete
pulsation cycle of the variable:  solutions using data of ascending and
descending branches of the light curve {\it alone}, for instance, lead to
identical radius and distance values for the Cepheid.

The data in Table 3 show that this is generally not true for the optical
solutions.  Here, the difference between the direct and the inverse
solution can be larger than 3$\sigma$ (although for most stars it is less
than this, especially for the longer-period Cepheids - see the data in
Table 3), so the choice of which fit to adopt is a matter of concern in the
optical solutions.  This choice has to be made for each star on the basis
of the quality and quantity of the observational data available for the
analysis, which determine the relative errors in the linear and angular
diameters, and taking into account the possible systematic errors in the
angular diameters introduced by microturbulence and gravity variations.
Therefore, the choice can be {\it different} for different stars.

Obviously, the insensitivity of the radii and distances derived with the
near-infrared methods to the choice of the way the fits are taken is one of
the great advantages of the infrared versions of the method over the
optical one.

\vspace{0.8cm}

\centerline {\bf 4.3. Phase Mismatch between Photometric and Radial Velocity
 Curves}

It has been known for a long time that the correct phase linking between the
radial velocity curve and the photometric curves in a BE analysis must be
done with high precision to avoid significant systematic errors.  We are
sensitive to this source of error because in most cases, the radial
velocity curve and the photometric curves of a given variable were not
measured contemporaneously, and the periods are not known with perfect
accuracy.

Figure 11 shows the typical dependence of the derived distance of a Cepheid
on the phase mismatch between the radial velocity and photometric curves we
find, in this case for the Cepheid U Sgr in the near-infrared K, J-K
solution.  Within a range of $\pm$ 0.05 in the phase mismatch, the typical
change in the radius and distance value of a Cepheid is about {\it 1\% per
phase mismatch of 0.01}.  Our method is therefore much less sensitive to
this phase mismatch than the maximum likelihood solutions of Laney \&
Stobie (1995) who find that their radius values change by a worrisome 5\%
for a phase mismatch of 0.01.  Figure 11 also shows the variation of the
relative accuracy of the radius and distance result with the misalignment
in phase between radial velocities and photometry; it is seen that there is
a fairly pronounced minimum in the error if the phase alignment is
correctly done.  We can therefore use this criterion to help establish the
correct phase relation between the radial velocity curve and the
photometric curves (although there may be a possibility that a part of
the observed phase shift is due to some intrinsic cause which we do not
yet fully understand).

Our experience shows that in all cases we can achieve the correct phase
alignment to within $\pm$ 0.01 using a) a visual alignment of the linear
displacement and angular diameter curves, and b) the criterion of a minimum
dispersion in the solution.  This is at least true for the {\it near-infrared
solutions}, due to the very low scatter in the angular diameters; for the
much noisier optical solutions, we might have failed to do the phase
alignment with this 1\% precision for some of the stars.  Also, the correct
phase alignment is hampered in the optical solutions by the systematic
differences in shape between the linear displacement and angular diameter
curves which are present for nearly all Cepheids, but which are not seen in
the infrared solutions (at least in the K, J-K solution; see section 4.4).
We can therefore state that for the infrared solutions, the systematic
errors we expect due to the phase alignment problem between the radial
velocities and the photometry are very likely to be below the 1\% level for
most of the stars.

\vspace{0.8cm}

\centerline {\bf 4.4.  Phase Mismatch between V and K Curves in the V,
 V-K Solution}

Another possible source of systematic error affecting the {\it V, V-K
solutions} is a possible misalignment of the V and K light curves which in
all instances have been obtained at different epochs, for the Cepheids in
our sample.  We must therefore check for the effect such a phase mismatch
between V and K has on the resulting radius and distance.  We did this by
introducing, as in the case of the velocity curves and photometry discussed
in the previous section, artificial phase shifts between the V and K light
curves, which leads to a {\it distortion of the resulting V-K color curve}
which is used in the radius and distance solution.  Our tests show that the
{\it change in the radius and distance is typically about 5\% for a phase
mismatch between V and K of 0.01}. This kind of phase mismatch (affecting
only the V, V-K solution) has therefore a much more serious effect on the
solutions than a possible phase mismatch between the radial velocities and
photometric curves discussed in the previous section.

As a consequence of an increasing phase mismatch between V and K, the {\it
amplitude} of the resulting V-K color curve, and the {\it phase of bluest
V-K color} change systematically.  We find that within a phase mismatch
interval from -0.05 to +0.05 the {\it amplitude ratio} A(V-K) / A(V)
decreases typically by 25\%, while at the same time the {\it phase shift
between the maxima of the V and V-K curves} varies from +0.08 to -0.04.
Since there seems to be little intrinsic variation from Cepheid to Cepheid
in both, the (V-K)/V amplitude ratio and the phase relation between the
maxima of both curves, the observed amplitude ratio and phase shift provide
in principle a useful criterion to deduce a phase mismatch between the V
and K curves, especially in cases where {\it both} values deviate from the
normal values in the expected sense.  However, due to the remaining
intrinsic spread in these quantities from Cepheid to Cepheid these criteria
are probably not too safe, and the only way to ensure a correct phase match
between V and K is to have very accurate periods, or ideally to use
contemporaneous or near-contemporaneous observations in both passbands.

As a conclusion, the correct phase alignment between the V and K curves is
critical in the method and must be done with great care.  The comparison of
the V, V-K radii and distances to the K, J-K solutions and the excellent
agreement found in section 3.1.  demonstrates, however, that we have
obviously succeeded to keep the errors from this source low, typically
below the 2\% level.  It is likely that most of the remaining scatter among
the two near-infrared solutions seen in Figs. 5 and 6 is due to small
errors in the phase alignment between the V and K curves of the variables.

\vspace{0.8cm}

\centerline {\bf 4.5. Uncertainties in the Calibrations of the Surface
 Brightness - Color Relations}
     
An obvious source of systematic uncertainty comes from possible errors in
the zero point and slopes of the three different surface brightness - color
relations from which the Cepheid angular diameters are deduced.  Our
discussion in Paper I shows that the zero point (common to the three
relations) is accurate to $\pm$0.003 while the uncertainties in the slope
values are different for the different calibrations.  We recall that the
zero point rests solely on interferometrically measured angular diameters
of cool giants and supergiants while the slopes were derived from a sample
of Cepheids using Thompson's method. We have further shown in paper I
that there is no detectable difference between the various surface
brightness-color relations defined by the giants and supergiants of our
calibrating sample. There is a possibility that stable supergiants have
somewhat different colors from the supergiants with pulsating atmospheres,
but the very good agreement of the slopes derived for the Cepheids with 
Thompson's method with those observed for the stable supergiants gives
us a lot of confidence that Cepheids are obeying surface brightness-color
relations which are basically indistinguishable from those obeyed by the
stable supergiants and giants.

Our tests show that in the infrared methods, a variation of the {\it zero
point} by 1$\sigma$ causes a change in the distance of 2.4\% and 2.2\%,
respectively, for the K, J-K and V, V-K versions of the method.  The radii do
not change because they do not depend on the zero point of the surface
brightness-color relation.  In the optical V, V-R method, the change in the
distance is very small, only 0.2\%.  A change in the respective {\it slope}
values by 1$\sigma$ causes changes in the distance and radius of 2.3\% and
2.8\% in the near-infrared K, J-K and V, V-K methods, respectively, while the
change in distance and radius in the optical method is 1.7\%.  These
results show that while these errors are still important for the
application of the different variants of the method, the effort we have made
in Paper I to reduce the uncertainties in the calibrations of the surface
brightness-color relations has been {\it crucial} to achieve the current
low level of systematic uncertainties in the method (see Table 5).

\newpage

\centerline {\bf 4.6.  The Conversion Factor from Radial to Pulsational
 Velocity}

In each of our solutions, both the radius and distance are directly
proportional to p, the factor adopted to convert the observed radial to
pulsational velocities of the stellar surface.  While most studies in the
past have used constant p-factors with values between 1.31 (Parsons 1972)
and 1.41 (Getting 1934), modern values over the last decade have clustered
near p=1.36.  The value to adopt has also a small dependence on the way the
radial velocities are measured.  An improvement came in the study of
Gieren, Barnes \& Moffett (1989) who introduced a period-dependent p-factor
(but constant for a given Cepheid) based on a formula which was derived
from the Cepheid models of Hindsley \& Bell (1989).  More recently, Sasselov
\& Karovska (1994) and Sabbey et al.  (1995) have advocated the use of a
p-factor which varies with phase during the pulsation of a Cepheid.  This
conclusion was based on non-LTE hydrodynamic models of Cepheid atmospheres
in conjunction with high-resolution optical and near-infrared spectra for a
few bright Cepheids.  However, the effect of a variable p-factor is not
seen in LTE calculations, and to date there remains some doubt about the
reality and the true size of this effect.

Our observational results in this paper, i.e.  {\it the near-perfect
agreement in the shapes of the linear displacement and angular diameter
curves} seen for nearly all Cepheids in the infrared solutions (especially
in the K, J-K solutions where we have not the phasing problem discussed in
section 4.4) {\it lend strong support to the conclusion that p=constant
must be a very good approximation for most Cepheids in our sample},
independent of their periods.  Otherwise, systematic deviations in the
shapes of the linear displacement curves vs.  the angular diameter curves
should show up, due to the distortions produced in the linear displacement
curve by the (wrong) assumption that p is constant over the pulsation
cycle.  There are a few Cepheids where such small deviations are indeed
observed; however, since we see them in the V, V-K solution, but {\it not}
in the K, J-K solution, their origin is almost certainly in a slight phase
mismatch between V and K, and not in neglecting a significant variation in
p over the pulsation cycle.

From a comparison of published modern values of the p-factor (constant for
a given Cepheid) we conclude that a realistic current uncertainty in the
radius and distance from any BE technique due to this factor is $\sim
\pm$2.5\%.  From all the possible errors affecting our infrared radii and
distances, the correct knowledge of p might be the largest remaining
uncertainty in the method.  Clearly, further work, both theoretical and
observational, will be very important to reduce this uncertainty from its
current level.

In Table 5, we summarize the discussion of section 4 and present a
corresponding error budget in the near-infrared {\it distance} to an
individual Cepheid.  The same errors apply to the radii except the
uncertainty in the zero point of the surface brightness-color relation.
{\it All possible systematic errors affecting a distance solution in the
infrared methods are smaller than 3\%}.  The total systematic error is
computed according to the precepts of Rabinovich (1995), assuming a uniform
distribution of each systematic error within the limits given in Table 5.
The total random error given in Table 5 reflects the random scatter in our
distance and radius solutions presented in Table 3 and refers to those
stars which have amplitudes $\geq$ 0.3 mag in the color curve used in the
solution, and very good observational data.  These random uncertainties can
be traced back to the uncertainties of the observations used in the
solutions. {\it For a typical large-amplitude Cepheid, we can easily reduce the
random error in its radius and distance to 1\% if the data used in the
analysis are excellent in quality and quantity}.  The total error is
obtained adding quadratically the total random and the total systematic
error.  It is found to be about 3\% for both near-infrared solutions.

It is worthwhile noting that over a {\it sample of Cepheids}, the errors
introduced by sources a) - d) in Table 5 tend to cancel out while errors e)
- h) are truly systematic in the sense that they do affect all the stars in
a sample in the same way.

Finally, we want to note that we do expect some sensitivity of the
calibrating relations to {\it metallicity}. It seems likely that the infrared
Barnes-Evans method has a lower sensitivity to metallicity than its optical
counterpart (whose metallicity sensitivity has been discussed by Gieren,
Barnes \& Moffett 1993) and does not introduce systematic errors larger
than the other ones described in this section, but this has yet to be
checked.

Our finding that the total systematic uncertainty of the infrared Cepheid
distances is about 3\% can be checked with independent methods of Cepheid
distance determination, and we will do this in the following section.

\vspace{1.5cm}

{\bf 5.  Comparison of the Infrared Distances to the ZAMS-Fitting Distances of} \\
\centerline {\bf the Parent Clusters and Associations}

The ZAMS-fitting distances to the 16 Cepheids of our sample have been reviewed
in Gieren \& Fouqu\'e (1993). We adopt the same distance moduli here, but we
try to estimate an associated uncertainty. Indeed, not all the ZAMS-fitting
distances have the same accuracy, depending on the photometric and
spectroscopic quality of the cluster/association data, the number of stars
involved in the fit, and the confidence in their selection (these stars must
be single, unevolved, main-sequence objects). Also systematic errors enter,
such as the degree of reliability of the association of the Cepheid with its
supposed parent cluster (which depends on the existence of radial velocity
information, the projected distance of the star onto the cluster, and is
obviously always more doubtful for OB associations than for well-defined 
clusters), the possible difference in metallicity with the template cluster
(Pleiades in all cases), and the true Pleiades distance modulus.

The resulting distances are given in Table 4, together with the ratio of
the ZAMS-fitting distance to our adopted distance, and the combined
uncertainty of this ratio.  The unweighted mean value of the ratio is $1.02
\pm 0.04$, which shows that both distance scales are in agreement.
However, some values appear to differ significantly from one.  In the
following, we discuss all cases where the ratio is smaller than 0.9 or
larger than 1.1:

{\bf SZ Tau}:  The ZAMS-fitting distance of NGC 1647 (Turner 1992) is of
 excellent quality, as well as our adopted distance for SZ Tau (from the
 $V$, $V - K$ solution).  But the star lies at $\sim 2^{\circ}$ from the
 cluster center, which corresponds to 19 pc at the distance of the cluster,
 and its proper motion disagrees with the mean proper motion of the cluster
 (Geffert \etal 1996).  We therefore prefer to exclude SZ Tau as a secure
 member of the NGC 1647 cluster.  We note that in Turner's study, SZ Tau is
 the only Cepheid which does not fit the nice period-luminosity relation,
 which is explained by Turner assuming that the star pulsates in the first
 overtone mode.  However, adopting our shorter distance would equally well
 put the star onto the relation.

{\bf CV Mon and T Mon}:  The ZAMS-fitting distance of the Mon OB2
 association (Turner 1976) is of better quality than the ZAMS-fitting
 distance of the anonymous cluster which is supposed to contain CV Mon
 (Turner 1978).  From Turner's work (1976, his Fig.  3), it appears that in
 fact the ``CV Mon'' cluster is at a compatible distance with the Mon OB2
 association.  Now, our adopted distance to CV Mon agrees with the
 ZAMS-fitting distance of Mon OB2.  We therefore adopt as the ZAMS-fitting
 distance of CV Mon the one for Mon OB2 in place of the less accurate
 ZAMS-fitting distance to the anonymous cluster to which it apparently
 belongs.  Our adopted distance to T Mon (from the $K$, $J - K$ solution)
 clearly disagrees with the ZAMS-fitting distance to Mon OB2.  We also note
 that Turner's arguments of membership are very weak.  Moreover, the
 reddening of T Mon ($E(B-V) = 0.209$) is clearly smaller than the
 reddening of CV Mon ($E(B-V) = 0.714$), which reinforces the conclusion
 that T Mon lies in the foreground of the Mon OB2 association.  We
 therefore reject T Mon as a member of the Mon OB2 association, and correct
 the ZAMS-fitting distance of CV Mon to the distance of Mon OB2, without
 excluding that it may also belong to the loose cluster discovered by van
 den Bergh (1957).

{\bf V340 Nor}: The ZAMS-fitting distance to the cluster NGC 6067 (Walker 1985)
 which contains both V340 Nor and QZ Nor is of excellent quality. This is not
 the case of our adopted distance to V340 Nor, because the $V$ light curve is
 poor (which makes the $V$, $V-K$ solution inaccurate) and the $J-K$ amplitude
 is small (making the $K$, $J-K$ solution inaccurate). However, V340 Nor only 
 lies (in projection) at 2' from the center of the cluster, and its radial 
 velocity (-40.34 $\pm$ 0.39 km s$^{-1}$ from Bersier \etal 1994) agrees very
 well with the mean radial velocity of the cluster ($-39.9 \pm 0.16$ 
 km~s$^{-1}$ from Mermilliod \etal 1987, from 8 stars). We therefore attribute 
 the discrepancy to the inaccuracy of the infrared Barnes-Evans distance, but 
 still consider V340 Nor as a rather secure member of NGC 6067. The other 
 Cepheid of the cluster, QZ Nor, has an adopted distance in good agreement with
 the ZAMS-fitting distance of the cluster, although it is located at 18' (in
 projection) from the cluster center, and has sometimes been rejected as a
 member of the cluster. In summary, we conclude that both stars are members of 
 NGC 6067, but we do not use the distance ratio of V340 Nor in the final 
 estimate of the agreement of both distance scales.

{\bf U Car}: This star together with VY Car are generally assigned to the 
 Car~OB2 association. Turner's (1988) study of this association does not give
 details about the quality of the ZAMS-fitting distance. Although the
 velocity measurements favour membership of both stars ($+1 \pm 2$ km~s$^{-1}$ 
 for Car OB2, $+2.0$ km~s$^{-1}$ for VY Car, and $+1.7$ km~s$^{-1}$ for U Car),
 we clearly find different distances to both Cepheids: the adopted distance to
 VY Car is in very good agreement with the ZAMS-fitting distance to Car OB2, 
 while the adopted distance to U Car puts it in the foreground of the 
 association. We therefore reject U Car as an association member.

{\bf SV Vul}: Turner (1974), quoted in Turner (1979), derived a ZAMS-fitting
 distance to the Vul OB1 association, which clearly disagrees with our rather
 accurate distance determination to SV Vul. We note that published reddening 
 measurements of the Cepheid span a large range ($E(B-V)$ values between 0.41 
 and 0.58), indicating that the region is complex. We consider that SV Vul is
 located in the background of the Vul OB1 association, although not as far as 
 the Vul OB2 association (which contains the Cepheid S Vul).

From this discussion, we conclude that four Cepheids in our sample do not 
belong to their supposed parent cluster or association (SZ Tau, T Mon, U Car
and SV Vul). Excluding these four stars as well as the uncertain distance ratio
of V340 Nor, and correcting the ZAMS-fitting distance to CV Mon, the mean ratio
(ZAMS-fitting distance to infrared Barnes-Evans distance) amounts to 
$0.987 \pm 0.016$ (unweighted), or $0.969 \pm 0.014$ (weighted). This shows 
that the agreement between both distance scales remain excellent after 
rejection of non-members. The standard deviation of the ratio is 0.05 
(unweighted: 0.053, weighted: 0.047), which means that both distance indicators
are accurate to better than 5\%, including systematic errors. For instance, a 
possible systematic error in the Pleiades distance modulus cannot exceed the 
published standard error ($5.57 \pm 0.08$ from van Leeuwen 1983). The exact 
accuracy of each method is difficult to estimate; however, the error budget 
presented in the previous section shows that the infrared surface brightness 
distances are accurate to 3\% in the mean, including random errors and 
identified systematic errors. We therefore attribute a 4\% accuracy in the mean
to the ZAMS-fitting distances.

\vspace{1.5cm}

\centerline {\bf 6.  Conclusions}

We have obtained the radii and distances of 16 supposed galactic cluster
Cepheids using our new optical and near-infrared calibrations of the
surface brightness method given in Paper I.  We confirm the conclusion of
Paper I that the radii and distances obtained from both infrared methods
(V, V-K and K, J-K) agree very well, and that the typical random uncertainty
of the infrared radii and distances is in the order of 2\%, while it is
generally much larger for the optical V, V-R solutions.  Our error budget
shows that the systematic errors affecting the infrared radii and distances
are in the order of 3\%, and that the total uncertainty of the radius and
distance of an individual Cepheid is $\sim$3\% provided there are very good
data and the amplitude of the color curve used in the surface brightness
solution exceeds $\sim$0.3 mag.  The optical solutions are generally
affected by both, large systematic and random errors, which are almost
certainly due to the variations in microturbulence and gravity during the
pulsation cycles of the Cepheids and which manifest themselves clearly in
the common plots of the linear and angular diameters versus phase.
However, there are a few stars where this problem is much less than for
others (like SW Vel, for instance), and it would be interesting to find out
the reason for this.  We speculate that the amount of systematic deviation
of the optical radius and distance from the corresponding infrared value is
correlated with the mean microturbulence of the Cepheid, which can vary
considerably from one Cepheid to another, even at an almost constant
period.  This hypothesis could be tested empirically by obtaining accurate
spectroscopic mean microturbulence values for the sample of variables
analyzed in this paper.

Our infrared radii agree very well with the infrared radii derived by Laney
\& Stobie with the maximum likelihood technique, providing further evidence
that both techniques are able to determine the radii of Cepheid variables
with an accuracy of $\sim$3\% (provided the amplitudes of the color and linear
displacement curves are not too small).

Our comparison of the adopted infrared distances of the variables to the
ZAMS-fitting distances of their supposed host clusters and associations
demonstrates that there is excellent agreement between both distance
scales.  Our accurate distance determinations, together with other
available information permit us to identify four non-members among the
present sample of Cepheids, while 2 stars (V Cen and BB Sgr) can now be
safely included into the list of cluster Cepheids.  From the comparison of
the infrared surface brightness to the ZAMS-fitting distances it also
follows that both methods yield distances accurate to better than 5
percent.  Adopting a 3 percent total uncertainty in our infrared distances,
as discussed in section 4, the mean accuracy of the ZAMS-fitting distance
determinations is about 4 percent, which is somewhat surprising given the
difficulties in most of the ZAMS-fitting distance determinations for the
present sample of Cepheids.  It would of course be interesting to include
the northern cluster Cepheids into the comparison, but since the present
sample includes $\sim$75 percent of the total sample of known or suspected
cluster Cepheids, we do not expect any significant change in our
conclusions reached in this paper. \\

Part of this work was completed while W.P.G.  was a Senior Visitor at the
European Southern Observatory.  He gratefully acknowledges the support
received during his stay.  W.P.G.  was also supported by Colciencias
through grant No.  190-91 to the Universidad Nacional de Colombia in
Bogot\'a, which is also gratefully acknowledged.  This work was completed
while W.P.G.  was on leave from the Observatorio Astron\'omico of the
Universidad Nacional de Colombia.  We have made use in this work of the very
useful compilation of galactic Cepheid data maintained by Donald Fernie at
David Dunlap Observatory (URL http://ddo.astro.utoronto.ca/cepheids.html),
and of the Cepheid photometric and radial velocity data archive maintained
by Douglas Welch at McMaster University (URL
http://www.\\physics.mcmaster.ca/Cepheid/).

\newpage
\centerline {\bf Appendix: Maximum Likelihood vs. Least-Squares Solutions}

There is a discussion in Laney \& Stobie (1995) about the respective merits
of the Maximum Likelihood and Least-Squares fits (hereafter ML and LS) to
determine the best values of the Cepheid radii (and distances, in our
case).  We fully agree with these authors that, {\it when uncertainties are
accurately known}, ML results are superior.  However, a small variation of
assumed uncertainties may lead to significant variations in the results of
the ML method.

In fact, the ML result always lies between the two extreme LS results,
namely the direct LS fit assuming no uncertainty in angular diameter values
(X axis) and the inverse LS fit, which assumes no uncertainty in linear
displacement values (Y axis).  Laney \& Stobie only compare the ML result
to the {\it direct} LS result, and find significant differences.  In fact, we
will show below that the ML fit is close to the {\it inverse} LS fit, because
relative accuracies are smaller in angular diameters than in linear
displacements.  So we agree with Laney \& Stobie to reject the direct LS
fit, but we prefer to adopt the inverse LS fit rather than the ML fit,
because the former has the desirable property of being independent of
assumed values of uncertainties.

In order to estimate which of the two variables carries more uncertainty, we
must compare the relative accuracy of measurements. As an estimator of linear 
displacement accuracy, Laney \& Stobie propose the ratio of $\sigma (r)$ 
(uncertainty of radial displacement) to $\Delta R$ (difference between maximum 
and minimum linear radius). The corresponding estimator of angular diameter 
accuracy would be the angular diameter accuracy ($\sigma (\phi) / \phi$), 
multiplied by some average angular diameter, and divided by the difference 
between maximum and minimum angular diameters. A more convenient, although
equivalent, choice of estimators consists of $\sigma (r) / R$, where $R$ is the
mean linear radius, and $\sigma (\phi) / \phi$.

In order to compute $\sigma (r)$ in solar radius units, we use the same
formula as Laney \& Stobie, first introduced by Balona (1977), namely:

$$\sigma (r) = 0.0844 \times P \times \frac{\sigma (RV)}{\sqrt N}
~~~~~~~~~~~~~~~~~~~~~~~~~~~~~~~~~~~~~~~~~~~~~~~~~~~~~~~~~~~~(A1)$$

\noindent where P is the Cepheid period in days, $\sigma (RV)$ is the mean
uncertainty of radial velocity measurements in km s$^{-1}$, and $N$ is the
number of radial velocity measurements.  The numerical factor 0.0844 is
$1.36 \times \frac{86400}{2 \times 696000}$.  Resulting values of $\sigma
(r) / R$ vary between 0.0016 (CV Mon) and 0.0083 (WZ Sgr), i.e.  they are
always better than 1\%.

Using Eqs.  1-3 of this paper, together with the adopted reddening ratios
of Paper I, we derive the following formulae to compute $\sigma (\phi) /
\phi$ for each surface brightness -- colour relation; we take into account
that $V$ and $K$ light curves are independent, but $V$ and $R_{\rm J}$, and
$J$ and $K$ are not: \\

\noindent
$ \frac{\sigma (\phi)}{\phi} (V, V - K)  = 
 \ln 10 \times \left[ \, 0.11 \  \sigma^2 (V) + 0.07 \ 
  \sigma^2 (K) + 0.011 \  \sigma^2 [E(B-V)] \right]~~~~~~~~~~~~~~(A2)$ \\
$ \frac{\sigma (\phi)}{\phi} (V, V - R_{\rm J}) = 
 \ln 10 \times \left[ \, 0.04 \  \sigma^2 (V) + 0.58 \ 
  \sigma^2 (V-R) + 0.007 \  \sigma^2 [E(B-V)] \right]~~~~~~~(A3)$ \\
$ \frac{\sigma (\phi)}{\phi} (K, J - K) = 
 \ln 10 \times \left[ \, 0.04 \  \sigma^2 (K) + 0.05 \ 
  \sigma^2 (J-K) + 0.002 \  \sigma^2 [E(B-V)] \right]~~~~~~~(A4)$ \\

For each Cepheid, an estimate of photometric uncertainties has been derived
from examination of light curves.  $\sigma (V)$ ranges from 0.01 to 0.015,
with extreme values of 0.02 (U~Car) and 0.03 (V340~Nor).  $\sigma (V-R)$
ranges from 0.01 to 0.02.  Note that the conversion from the Cousins to
Johnson system does not affect the accuracy of the measurements, but may
only introduce a systematic error in the derived radius and distance.
$\sigma (K)$ and $\sigma (J-K)$ are taken as 0.01, except for EV~Sct where
infrared light curves are not well sampled, and we estimate both
uncertainties to be 0.03.

To estimate $\sigma [E(B-V)]$, we have used Fernie's database of reddenings 
(Fernie \etal 1995), which gives several measures for each star, to which we 
added Bersier's (1996) measurements, when available (5 stars). This gives 
between 7 and 11 values per star, from which we derive an estimate of 
$\sigma [E(B-V)]$. Resulting values range from 0.025 (S~Nor) to 0.072 (SV~Vul).
For 2 stars, few available values lead to uncertain results: V340~Nor 
(3 values, $\sigma [E(B-V)] = 0.017$) and QZ~Nor (4 values, 
$\sigma [E(B-V)] = 0.050$).

From these individual uncertainties, we compute the relative accuracy of
angular diameter measurements for each colour -- magnitude combination.
For $V$, $V - R_{\rm J}$, results range from 0.019 to 0.038, and the ratios
to the corresponding relative accuracy of linear displacements range from 3
to 15.  For $V$, $V - K$, relative accuracies range from 0.011 to 0.024,
and ratios from 2 to 13.  And for $K$, $J - K$, relative accuracies range
from 0.007 to 0.021, and ratios from 1 to 11.  {\it Therefore, if we only
consider measuring uncertainties, we are already justified to use the
inverse least-squares solution in all cases}.  Now, looking back at the
method used to derive angular diameters, any contribution of gravity or
microturbulence differences among stars adds to the uncertainty of the
angular diameter determination.  This is clearly evidenced in the $V$, $V -
R_{\rm J}$ solution, where the measuring uncertainty alone does not explain
the large dispersion observed in the linear displacement vs.  angular
diameter diagram.  This reinforces our decision to use the {\it inverse} LS
solution as the best choice.

Finally, let us say that in cases where this decision may be disputed 
(typically when the ratio of the relative accuracies is smaller than 2, and 
assuming that systematic uncertainties in the angular diameter determination 
may be neglected), the difference between direct and inverse LS solutions 
(as given in Table 3, columns 4 and 6) is generally smaller than the mean error
of the adopted solution, which makes the result insensitive to our choice.

\newpage
\centerline {\bf References}

\begin{itemize}
\item[-] Balona, L.A. 1977, MNRAS, 178, 231
\item[-] Barnes, T.G., Moffett, T.J., \& Slovak, M.H. 1988, ApJS, 66, 43
\item[-] Berdnikov, L.N. 1986, Variable Stars, 22, No. 3, 369
\item[-] Berdnikov, L.N. 1987, Variable Stars, 22, No.4, 530
\item[-] Berdnikov, L.N., \& Turner, D.G. 1995, Astronomy Letters, Vol. 21,
 No.6, 717
\item[-] Bersier, D. 1996, A\&A, in press
\item[-] Bersier, D., Burki, G., Mayor, M., \& Duquennoy, A. 1994, A\&AS, 108, 25
\item[-] Breger, M. 1970, AJ, 75, 239
\item[-] Coulson, I.M. 1983, MNRAS, 203, 925
\item[-] Coulson, I.M., \& Caldwell, J.A.R. 1985a, SAAO Circ., 9, 5
\item[-] Coulson, I.M., \& Caldwell, J.A.R. 1985b, MNRAS, 216, 671
\item[-] Eggen, O.J. 1983, AJ, 88, 379
\item[-] Evans, N.R., \& Udalski, A. 1994, AJ, 108, 653
\item[-] Feast, M.W., \& Walker, A.R. 1987, ARA\&A, 25, 345
\item[-] Fernie, J.D. 1990, ApJS, 72, 153
\item[-] Fernie, J.D., Beattie, B., Evans, N.R., \& Seager, S. 1995, IBVS,
 No. 4148  
\item[-] Ferrarese, L., et al. 1996, ApJ, 464, 568
\item[-] Fouqu\'e, P., \& Gieren, W.P. 1997, A\&A, in press (Paper I)
\item[-] Getting, I.A. 1934, MNRAS, 95, 141
\item[-] Geffert, M., Bonnefond, P., Maintz, G., \& Guibert, J. 1996, A\&AS,
 118, 277
\item[-] Gieren, W.P. 1981a, ApJS, 46, 287
\item[-] Gieren, W.P. 1981b, ApJS, 47, 315
\item[-] Gieren, W.P. 1984, ApJ, 282, 650
\item[-] Gieren, W.P. 1989, A\&A, 216, 135
\item[-] Gieren, W.P., Barnes, T.G., \& Moffett, T.J. 1989, ApJ, 342, 467
\item[-] Gieren, W.P., Barnes, T.G., \& Moffett, T.J. 1993, ApJ, 418, 135
\item[-] Gieren, W.P., \& Fouqu\'e, P. 1993, AJ, 106, 734
\item[-] Gieren, W.P., Mermilliod, J.C., Matthews, J.M., \& Welch, D.S. 1996,
 AJ, 111, 2059
\item[-] Hindsley, R.B., \& Bell, R.A. 1989, ApJ, 341, 1004
\item[-] Laney, C.D., \& Stobie, R.S. 1992, A\&AS, 93, 93
\item[-] Laney, C.D., \& Stobie, R.S. 1995, MNRAS, 274, 337
\item[-] Mermilliod, J.C., Mayor, M., \& Burki, G. 1987, A\&AS, 70, 389
\item[-] Metzger, M.R., Caldwell, J.A.R., McCarthy, J.K., \& Schechter, P.L.
 1991, ApJS, 76, 803
\item[-] Metzger, M.R., Caldwell, J.A.R., \& Schechter, P.L. 1992, AJ, 103, 529
\item[-] Moffett, T.J., \& Barnes, T.J. 1984, ApJS, 55, 389
\item[-] Parsons, S.B. 1972, ApJ, 174, 57
\item[-] Rabinovich, S. 1995, Measurement Errors: Theory and Practice (American
 Institute of Physics: New York)
\item[-] Sabbey, C.N., Sasselov, D.D., Fieldus, M.S., Lester, J.B., Venn, K.A.,
 \& Butler, R.P. 1995, ApJ, 446, 250
\item[-] Saha, A., Sandage, A., Labhardt, L., Tammann, G.A., Macchetto, F.D.,
 \& Panagia, N. 1996, ApJ, 466, 55
\item[-] Sasselov, D.D., \& Karovska, M. 1994, ApJ, 432, 367
\item[-] Szabados, L. 1977, Mitt. Konkoly Observatory, No. 70.
\item[-] Turner, D.G. 1974, Ph.D. Thesis, Univ. Western Ontario, London, Ontario, Canada
\item[-] Turner, D.G. 1976, ApJ, 210, 65
\item[-] Turner, D.G. 1978, JRASC, 72, 248
\item[-] Turner, D.G. 1979, JRASC, 73, 74
\item[-] Turner, D.G. 1988, ASP Conf. Ser., 4, 178
\item[-] Turner, D.G. 1992, AJ, 104, 1865
\item[-] van den Bergh, S. 1957, ApJ, 126, 323
\item[-] van Leeuwen, F. 1983, Ph.D. Thesis, Leiden University, The Netherlands
\item[-] Walker, A.R. 1985, MNRAS, 214, 45
\item[-] Walraven, J.H., Tinbergen, J. \& Walraven, T. 1964, BAN, 17, 520
\item[-] Welch, D.L. 1994, AJ, 108, 1421
\end{itemize}

\newpage
\centerline {\bf Figure Captions}

\begin{description} 

\item[Fig.  1:] {\it Top panel}:  The angular diameters
 (open circles) of the Cepheid VY Car calculated from the optical V, V-R
 photometry, plotted against phase.  Overplotted (dots) is the linear
 displacement variation calculated from the integrated radial velocity
 curve of the Cepheid.  {\it Bottom panel}:  Plot of the linear
 displacements versus the angular diameters.  Overplotted is the
 least-squares inverse fit which takes into account that the angular
 diameters are more uncertain than the linear displacements.

\item[Fig.  2:] Same as Fig.  1, for the angular diameters calculated from
 the infrared K, J-K magnitude-color combination (equ. 3; see text).  Note
 the marked increase in accuracy of the solution, as compared to the
 optical solution shown in Fig. 1.

\item[Fig.  3:] Same as Fig. 1 and 2, for the angular diameters calculated
 from the V, V-K magnitude-color combination.  The accuracy of the solution
 is comparable to the one of the K, J-K solution.

\item[Fig.  4:] The uncertainties of the radius and distance solutions
 (normalized to a common number of 30 photometric observations), plotted
 against the amplitude of the color curve used in the solution, for the
 three magnitude-color combinations used in this work.  The V-K amplitude
 is large enough even for the lowest-amplitude Cepheids not to affect the
 accuracy of the solution (bottom panel).  In the K, J-K solutions, there
 is a sharp increase in the uncertainty if the J-K amplitude drops below
 0.1 mag.  For amplitudes larger than about 0.3 mag, the random uncertainty
 can be reduced to $\sim$1 percent in both infrared solutions.

\item[Fig.  5:]  The ratio of the radii obtained from the V, V-K version of
 the method to those obtained from K, J-K, plotted against the pulsation
 period.  The mean ratio is very close to unity, independent of period.
 Note the very low scatter in the diagram.The K, J-K solutions for the
 three shortest-period Cepheids (open squares) are unreliable due to the
 very low J-K amplitudes.

\item[Fig. 6:]  The same as Fig. 5, for the distances. 

\item[Fig.  7:]  The angular diameters (open circles) and linear
 displacements (dots) plotted against phase, for the Cepheid U Car and for
 the K, J-K (top panel) and V, V-K solution (bottom panel).  Note the
 small, but systematic deviation of the angular diameters from the linear
 displacements in the V, V-K solution near phase 0.3 which is not seen in
 the K, J-K solution.  This effect is very likely due to a slight phase
 mismatch between the V and K light curves used in the V, V-K solution.

\item[Fig.  8:]  The ratio of the radii obtained from the optical V, V-R
 solution to the adopted infrared radii (see text), plotted against
 pulsation period.  Note the large scatter, as compared to Fig.  5.  Also,
 the mean ratio is clearly larger than unity.

\item[Fig. 9:]  Same as Fig. 8, for the distances. Compare to Fig. 6.

\item[Fig.  10:]  Comparison of the maximum likelihood infrared radii of
 Laney \& Stobie (1995) to our infrared solutions.  The three
 shortest-period and lowest-amplitude Cepheids (open squares) have
 unreliable maximum likelihood radii.  For the remainder of the Cepheids,
 the agreement of the radii is very good.  The mean radius ratio is very
 close to unity, and the scatter in the diagram is very low.

\item[Fig.  11:]  Variation of the distance for the Cepheid U Sgr
 calculated from our K, J-K method as a function of the phase mismatch
 between the photometric and the radial velocity curve (solid curve).  The
 change in the distance is $\sim$1 percent for a change in the phase alignment
 of 0.01.  Also plotted (dotted curve) is the variation of the relative
 distance uncertainty with the phase mismatch.  The minimum of this curve
 can be used as a criterion to determine the correct phase relation between
 radial velocities and photometry.

\end{description}
\end{document}